\documentclass[aps,pre,preprint,superscriptaddress]{revtex4-1}
\usepackage{amsmath}
\usepackage{amssymb}
\usepackage{wasysym}
\usepackage{xcolor}
\usepackage{graphicx}
\usepackage{endnotes}
\begin{document}

\title{Diffusion, mixing, and segregation in confined granular flows}
\author{Alexander M. Fry}
\affiliation{Department of Mechanical Engineering, Northwestern University, Evanston, IL 60208, USA}
\author{Paul B. Umbanhowar}
\affiliation{Department of Mechanical Engineering, Northwestern University, Evanston, IL 60208, USA}
\author{Julio M. Ottino}
\affiliation{Department of Mechanical Engineering, Northwestern University, Evanston, IL 60208, USA}
\affiliation{Department of Chemical and Biological Engineering, Northwestern University, Evanston, IL 60208, USA}
\affiliation{Northwestern Institute on Complex Systems (NICO), Northwestern University, Evanston, IL 60208, USA}
\author{Richard M. Lueptow}
\affiliation{Department of Mechanical Engineering, Northwestern University, Evanston, IL 60208, USA}
\affiliation{Department of Chemical and Biological Engineering, Northwestern University, Evanston, IL 60208, USA}
\affiliation{Northwestern Institute on Complex Systems (NICO), Northwestern University, Evanston, IL 60208, USA}
\date{\today}

\begin{abstract}
Discrete element method simulations of confined bidisperse granular shear flows elucidate the balance between diffusion and segregation that can lead to either mixed or segregated states, depending on confining pressure. Results indicate that the collisional diffusion is essentially independent of overburden pressure. Because the rate of segregation diminishes with overburden pressure, the tendency for particles to segregate weakens relative to the re-mixing of particles due to collisional diffusion as the overburden pressure increases. Using a continuum approach that includes a pressure dependent segregation velocity and a pressure independent diffusion coefficient, the interplay between diffusion and segregation is accurately predicted for both size and density bidisperse mixtures over a wide range of flow conditions when compared to simulation results. Additional simulations with initially segregated conditions demonstrate that applying a high enough overburden pressure can suppress segregation to the point that collisional diffusion mixes the segregated particles.

\end{abstract}

\maketitle

\section{Introduction\label{Introduction}}
Flowing mixtures of particles varying in size or density rearrange, leading to either mixing or segregation (de-mixing) depending on whether the dominant species-specific motions are random or directed~\cite{savageLun1988,ottinoKhakhar2000,dolgunin1995,grayThornton2005,fan2014}. The random collision-driven re-arrangements are understood to be diffusion-like, with the diffusion coefficient depending on the product of the shear rate and the square of the mean particle diameter~\cite{bridgwater1980,utter2004}. The species-specific directed motions that lead to segregation are functions of particle properties, such as the size ratio~\cite{savageLun1988,ottinoKhakhar2000,drahunBridgwater1983,schlick2015} or density ratio~\cite{tripathi2013,xiao2016,liu2017}, and flow conditions, including the shear rate and local particle concentration~\cite{savageLun1988,dolgunin1995,fan2014,gray2006}. 

In contrast to free surface flows, the influence of confining pressure on segregation and diffusion in granular shear flows has been relatively unexplored. For segregation, experimental studies using split-bottom or annular shear cells under variable confining pressure indicate that the rate of segregation in granular shear flows decreases with increasing overburden~\cite{hillFan2008,golickDaniels2009}. For diffusion, on the other hand, the confining pressure does not significantly affect diffusion of mm-sized glass beads in an experimental simple shear experiment~\cite{bridgwater1980,bridgwater1994}. Likewise, in DEM simulations of low concentrations of small particles in shear flows of large particles, the diffusion is only a weak function of overburden pressure at small overburden pressures~\cite{khola2016}.

We recently used DEM simulations to show that in dense flows of equal volume mixtures of size or density bidisperse particles, the rate and ultimate degree of segregation are decreasing functions of the overburden pressure~\cite{fry2018}. Furthermore, we found that the segregation rate in size or density bidisperse flows is nearly linearly dependent over a wide range of flow conditions on the inertial number, $I=\dot{\gamma} d / \sqrt{P/ \rho}$, which includes the effects of shear rate, $\dot{\gamma}$, particle diameter, $d$, overburden pressure, $P$, and particle density, $\rho$. A similar inertial number dependence was found for the rate of density segregation of a single intruder particle in shear flows of less dense particles~\cite{liu2017}. 

To extend the understanding of overburden pressure effects on diffusion and segregation sufficiently to predict the state of mixedness in confined granular flows, we measure the diffusion coefficient in steady granular shear flow under various flow and confinement conditions. Using this understanding of the pressure dependence of diffusion, combined with the previously measured segregation dependence on pressure, we are able to apply a continuum model for segregation, similar to that described in previous research~\cite{dolgunin1995,gray2006,grayAncey2015,mayDaniels2010,fan2014,xiao2016}. At overburdens much larger than the lithostatic pressure at the bottom of the particle bed, i.e., $P\gg \rho_B g h$, where $\rho_B$ is the bulk density of the particles, $g$ is the acceleration due to gravity, and $h$ is the depth of the particle bed, the continuum model predicts that the system reaches a near fully-mixed steady state, irrespective of whether the initial conditions are mixed or segregated. This suggests that a simple shear cell under high confining pressure can mix size or density bidisperse particles at shear rates that would normally lead to near-complete segregation without an imposed overburden, such as in free-surface flows. To test this concept, we perform a series of simple shear simulations under high confining pressures beginning from initially segregated conditions, to test whether overburden pressure can be used to drive mixing for materials that would normally segregate in free surface flows.

 
\section{Variable overburden shear flow\label{simulationScheme}}
\begin{figure}[t]
\begin{center}
\includegraphics[width=0.475\columnwidth]{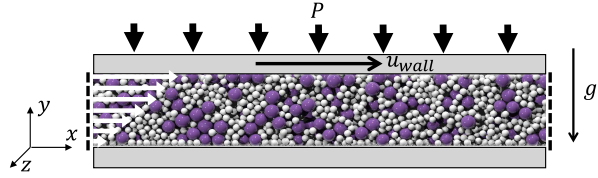}
\caption{Schematic (side view) of size bidisperse (diameter $d_{S}=2 \, \mathrm{mm}$ and $d_{L}=4 \, \mathrm{mm}$) shear flow $0.5 \, \mathrm{s}$ after the onset of shear (before significant segregation occurs). Vertical dashed lines represent streamwise periodic boundary conditions; the domain is also bounded by periodic boundary conditions in the spanwise direction. Top and bottom walls are modeled as flat frictional planes. Top wall mass is varied to change the overburden pressure $P$, and the top wall is free to move vertically due to dilation or compaction of the particle bed. The particle bed shown here is truncated in the streamwise direction compared to actual simulations. Source: Reproduced with permission from Fry et al.~\cite{fry2018}. Copyright 2018 the American Physical Society.}\label{schematic}
\end{center}
\end{figure}

The effect of confining pressure (overburden) on diffusion is studied here using DEM simulations described in detail in a previous study~\cite{fry2018} and here in the Appendix. Briefly, a massive top wall is placed on a bed of size or density bidisperse mm-sized spherical particles and is moved horizontally at constant velocity to generate a simple planar shear flow under a specific confining pressure, $P$, as shown schematically in Fig.~\ref{schematic}. The top wall is free to move vertically to provide a constant overburden, and exhibits small temporal fluctuations in height ($<2\%$) as the bed dilates to accommodate different packings of the bed particles. The distance between the top and bottom walls is $h \approx 25 \, \mathrm{mm}$, the length of the domain in the streamwise direction is $l = 280 \, \mathrm{mm}$, and the width in the spanwise direction is $w=35 \, \mathrm{mm}$. The domain is periodic in the spanwise and streamwise directions, i.e.\ particles leaving from one side of the domain re-enter on the opposite side. In order to avoid changes in the velocity profile with changing overburden pressure, a linear with depth velocity stabilizing force is applied to each particle in the streamwise direction at every time step, according to
\begin{equation}\label{velocityControlScheme}
F_{stabilize}=A(\dot{\gamma}y-u_{x}),
\end{equation}
 where $u_{x}$ is the particle's streamwise velocity, $y$ is the particle's vertical position, $\dot{\gamma}$ is the imposed global shear rate, and $A$ is a control parameter~\cite{fry2018}. A similar stabilizing force scheme has been used to control the velocity profile in simulations of frictionless particles~\cite{lerner2012,clark2018}. Since the normal and frictional forces that tend to alter the linear velocity profile scale with the local pressure, $A$ is varied proportionally to the local pressure in the bed. In this way, the shear rate is nearly constant throughout the depth and under all confinement conditions, which allows direct study of the effects of overburden on diffusion as in previous work on segregation~\cite{fry2018}.
 
Simulations are performed for both size bidisperse (large to small particle diameter ratio $d_{L}/d_{S}=2$) and density bidisperse mixtures (heavy to light particle density ratio $\rho_{H}/\rho_{L}=9$) under a wide range of pressures ($85 \, \mathrm{Pa} \leq P \leq 21 \, \mathrm{kPa}$) and several mean particle sizes [$\bar{d}=(d_L+d_S)/2$], mean particle densities [$\bar{\rho}=(\rho_H+\rho_L)/2$], shear rates, and gravitational accelerations (see Table~\ref{simulationConditions}). Particle species of a given nominal size have a uniformly distributed size polydispersity of $\pm10\%$ in size bidisperse cases and $\pm20\%$ in density bidisperse cases to reduce particle layering or other ordering. The number of particles in each simulation ranges from $10^4$ to $10^5$, depending on the particle diameter.

\begin{table}[ht]
\caption{Simulation conditions for data in Figs.~\ref{diffusion1} --~\ref{continuumAndDEM_allCases}}\label{simulationConditions}
\begin{tabular}{lccccc}
\hline
\hline
  Symbol & $\bar{d} \, \mathrm{[mm]}$ & $\bar{\rho} \, \mathrm{[kg/m^3]}$ & $\dot{\gamma} \, \mathrm{[s^{-1}]}$ & $g \, \mathrm{[m/s^{2}]}$ &\\[1pt]
    \hline
       $\textcolor{blue} \blacktriangledown$ & 3 & 2500 & 2.5 & 9.8 &\\
       $\textcolor{black} \Circle$ &3 & 2500 & 5 & 9.8 &\\
       $\textcolor{blue} \blacktriangleright$ & 3 & 2500& 10 & 9.8 &\\
       $\textcolor{blue} \blacktriangle$ & 3 & 2500 & 25 & 9.8 &\\
       $\textcolor{green} \bigstar$  & 3 & 2500 & 5 & 19.6 &\\
       $\textcolor{red} \blacklozenge$ & 3 & 1250 & 5 & 9.8 &\\
       $\textcolor{red} \blacksquare$ & 3 & 5000 & 5 & 9.8 &\\
       $\textcolor{cyan} \CIRCLE$ & 1.5 & 5000 & 5 & 9.8 &\\
         \hline
         \hline
  \end{tabular}
\end{table}

\section{Influence of Overburden on Diffusion\label{Relationship between overburden and diffusion}}

To quantify particle diffusion, we measure the mean squared displacement of every particle in the spanwise ($z$) direction, and average it over the system at time $t$ as
\begin{equation}
MSD_{z}(\Delta t)= \frac{1}{n_p} \sum\limits_{n=1}^{n_p} [z_n(t+\Delta t) - z_n(t)]^2,
\end{equation}
where $n_p$ is the number of particles in the system. The diffusion coefficient $D_{z}$ is then fit to the data as
\begin{equation}
D_{z}=MSD_{z}(\Delta t)/(2\Delta t),
\end{equation}
where $\Delta t$ is a time interval over which the particle motion is diffusive. The $MSD_z$ data (not shown) is analyzed in each simulation over the time interval $0.1 \, \mathrm{s} \leq t \leq 3 \, \mathrm{s}$ after the onset of shear as segregation is just beginning, and is linear as a function of time, indicating that the behavior is diffusive. We consider the spanwise diffusion, since it is more difficult to measure in other directions due to the mean flow in the streamwise ($x$) direction and the segregation in the vertical ($y$) direction. Furthermore, since the boundary conditions are periodic in the spanwise direction, the confinement effects that occur in the vertical direction, which is bounded by walls on the top and bottom, are avoided. Nevertheless, measurements of $D_{y}$ (not shown) indicate that its magnitude and pressure-dependence are very similar to the spanwise diffusion coefficient when segregation and the influence of bounding walls are accounted for. In contrast, after accounting for streamwise advection, measurements of $D_{x}$ indicate that it is super-diffusive. We note that the diffusion tensor has been shown to be isotropic in some studies~\cite{bridgwater1994,savage1993} but not in others~\cite{campbell1997,hsiau1999,utter2004}. In any case, since segregation in this planar shear flow geometry is solely in the vertical direction, the influence of streamwise diffusion is not a factor in the following analyses. 

\begin{figure}[t]
\begin{center}
\includegraphics[width=0.375\columnwidth]{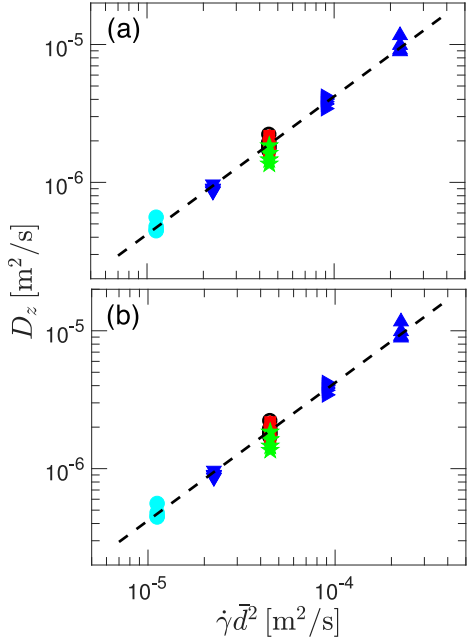}
\caption{Spanwise diffusion coefficient, $D_{z}$, vs.\ $\dot{\gamma}\bar{d}^2$ for (a) size bidisperse and (b) density bidisperse flows. Data are for the various flow conditions described in Table~\ref{simulationConditions} and spanning a range of overburden pressures $85 \, \mathrm{Pa} \leq P \leq 21 \, \mathrm{kPa}$, for a total of 43 size bidisperse cases in (a) and 44 density bidisperse cases in (b). Dashed lines (- -) are the correlation $D_{z}=C_{D} \dot{\gamma} \bar{d}^2$, with $C_{D}=0.042$ for both size and density bidisperse flows.}\label{diffusion1}
\end{center}
\end{figure}

Previous studies~\cite{bridgwater1980,utter2004,fan2013} report a linear relationship between the diffusion coefficient and the product of the shear rate and the square of the mean particle diameter. To test this relation, $D_z$ is plotted versus $\dot{\gamma} \bar{d}^2$ in Fig.~\ref{diffusion1}(a,b). The data collapse onto the same line, with a slope of 1, for both size and density bidisperse flows under a range of pressure and flow conditions, i.e.\ $D_{z}= C_{D} \dot{\gamma} \bar{d}^2$. The leading coefficient is nearly identical for both the size bidisperse and density bidisperse cases, $C_{D,size}=0.0427 \approx C_{D,density}=0.0413$, indicating that the dependence of $C_{D}$ on size or density dispersity is negligible. The coefficient $C_{D}=0.042$ will be used in the segregation continuum modeling presented in Sec.~\ref{continuumModel} for both size and density bidisperse flows. This value of $C_{D}$ is comparable with previous reported values for the transverse (non-flow) directions ($C_{D,y}$ and $C_{D,z}$) for dense granular flows, which have been reported experimentally by Bridgwater (0.051-0.057~\cite{bridgwater1980}), Hsiau and Shieh (0.020-0.025~\cite{hsiau1999}), and Utter and Behringer (0.108~\cite{utter2004}) and computationally by Savage and Dai (0.051~\cite{savage1993}) and Fan et al. (0.01-0.1~\cite{fan2014,fan2015}).

\begin{figure}[t]
\begin{center}
\includegraphics[width=0.425\columnwidth]{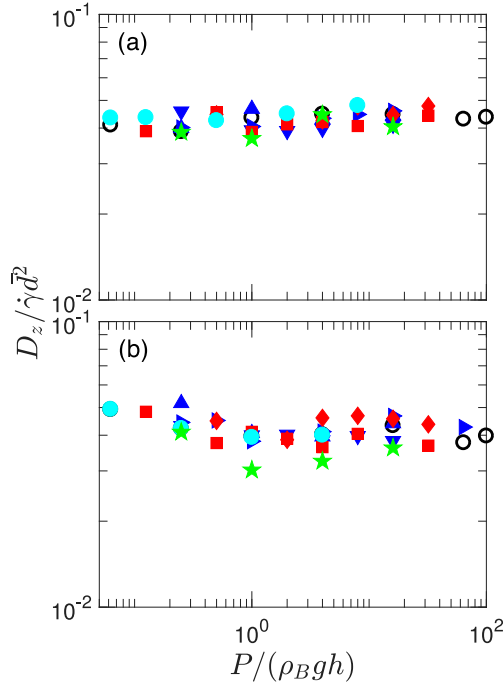}
\caption{Normalized spanwise diffusion coefficient, $D_{z}/\dot{\gamma}\bar{d}^{2}$ vs.\ normalized overburden pressure, $P/(\rho_B g h)$, for (a) size bidisperse and (b) density bidisperse flows. Data points are plotted for the various flow conditions described in Table.~\ref{simulationConditions}.}\label{diffusion2}
\end{center}
\end{figure}

To demonstrate the effect of overburden pressure on diffusion, the normalized diffusion coefficient, $D_{z}/ (\dot{\gamma} \bar{d}^2)$ is plotted versus\ normalized overburden $P/(\rho_B g h)$ in Fig.~\ref{diffusion2} for size and density bidisperse data from shear flows with varying $P$, $\dot{\gamma}$, $\bar{d}$, $\bar{\rho}$, and $g$. The figure shows that diffusion is independent of pressure for both size and density bidisperse flows over a range of overburden pressures that corresponds to two orders of magnitude in the mean pressure in the bed. This is consistent with previous experimental~\cite{bridgwater1980} and computational~\cite{khola2016} results, which both indicate a small or negligible dependence of diffusion on pressure, though at much smaller overburden pressures. The data also show that $D_{z}$ is independent of gravitational acceleration for the two values investigated ($9.8 \, \mathrm{m/s^2}$ and $19.6 \, \mathrm{m/s^2}$). It is surprising that the diffusion coefficient, normalized by $\dot{\gamma}\bar{d}^2$, is nearly identical for cases under such a wide range of flow and pressure conditions and also in both size bidisperse and density bidisperse flows. The implication is that diffusion is simply a geometry-dependent random walk, influenced by the nature of granular packings and possibly particle shape, while the segregation -- which depends dramatically on overburden pressure -- appears to be force-dependent.

\section{Continuum modeling of confined flows}\label{continuumModel}
Having determined the pressure dependence of segregation previously~\cite{fry2018} and here diffusion in size and density bidisperse flows, we test whether the continuum model developed previously for free surface flows can accurately predict the mixing behavior of these confined flows. This transport equation for granular flows~\cite{dolgunin1995,gray2006,grayAncey2015,mayDaniels2010,fan2014,xiao2016}, adds a species-specific segregation flux to the advection-diffusion equation:
\begin{equation}\label{s-a-d_equation}
\frac{\partial c_{i}}{\partial t} + \nabla \cdot (\mathbf{u} c_{i})+ \frac{\partial}{\partial y}(w_{p,i} c_{i})= \nabla \cdot (D\nabla c_{i}),
\end{equation}
where $c_i$ is the concentration of a given species, $\mathbf{u}$ is the mean velocity field, $D$ is the diffusion coefficient, which is assumed to be isotropic~\cite{bridgwater1994,fan2014}, and $w_{p,i}=S \dot{\gamma}(1-c_i)$ is the species-specific segregation velocity, with $S$ a coefficient dependent on particle size ratio~\cite{schlick2015} or density ratio~\cite{xiao2016}. This expression for $w_{p,i}$ is consistent with the kinetic sieving model of Savage and Lun~\cite{savageLun1988} for size bidisperse particles. Although a relation that is nonlinear in concentration~\cite{vanderVaart2015,jones2018} describes the segregation flux slightly more accurately over the full range of concentrations, the first order relation for $w_{p,i}$ above, which is linear in $c$, is sufficiently accurate for the purposes of this paper.

For simplicity, we compare the steady state solution of the continuum equation to the steady state segregation from the DEM results. Since the conditions are steady and fully developed, and since there is no net velocity in the vertical or spanwise directions, the first two terms of equation (\ref{s-a-d_equation}) are zero, and the right hand side depends only on the second derivative in $y$. The resulting ordinary differential equation has a solution for the predicted depthwise concentration profiles with the simple form~\cite{gray2006,schlick2015b}:
\begin{equation}\label{steadyStateContinuumEquation}
c_i(\tilde{y})=\frac{1}{1+Ae^{-\tilde{y}/\lambda}},
\end{equation}
where $\lambda=C_{D} \bar{d}^2/(S h)$ is a nondimensional ratio of segregation to diffusive time scales, $C_{D}=0.042$ is the leading coefficient for diffusion described in Fig.~\ref{diffusion1}, $S$ is the segregation coefficient in the segregation velocity equation, $h$ is the height of the shear cell, and $A=e^{1/(2\lambda)}$ is an integration constant for equal volume mixtures that satisfies the condition $\int_{0}^{1}c_i(\tilde{y})d\tilde{y}=0.5$. Note that the steady-state degree of segregation is insensitive to the shear rate, since both $D$ and $w_{p,i}$ are linear in $\dot{\gamma}$~\cite{schlick2015b}. Note also that the steady state degree of segregation increases with the depth of the bed, as the relative size of the layers adjacent to the ``mixing interface" at $\tilde{y}=0.5$ decreases compared to the total depth of the bed, i.e. in deeper beds particles must diffuse a longer distance to impact the degree of segregation than in shallow beds. This explains the finding of the previous study \cite{fry2018} that the bed height $h$ is a relevant length scale for the steady-state degree of segregation but not for the rate of segregation (where the relevant length scale is the mean particle size $\bar{d}$). 

The segregation coefficient, $S$ is calculated by combining a size ratio~\cite{schlick2015} or density ratio~\cite{xiao2016} dependent correlation with a correction for overburden pressure~\cite{fry2018} determined previously for size and density bidisperse flows, respectively, as
\begin{equation}\label{segregationCoefficientSize}
S_{size}= 0.26 d_{S} \ln{(d_L/d_S)} \sqrt{\bar{\rho}g\bar{d}/\bar{P}}
\end{equation}
\begin{equation}\label{segregationCoefficientDensity}
S_{density}= 0.081 \bar{d} \ln{(\rho_H/\rho_L)} \sqrt{\bar{\rho}g\bar{d}/\bar{P}},
\end{equation}
where $d_{S}$ and $d_L$ are the small and large particle diameters, respectively, in size bidisperse mixtures, $\bar{d}$ is the mean particle diameter [$\bar{d}=(d_{S}+d_{L})/2$ for size bidisperse mixtures and $\bar{d}=d_{H}=d_{L}$ for density bidisperse mixtures], and $\rho_{H}$ and $\rho_{L}$ are the heavy and light particle densities, respectively, in density bidisperse mixtures. In order to apply the analytical solution, we calculate the segregation velocity for each simulation based on the value of the pressure at the midline of the shear cell, $\bar{P}=P+\rho_{B} g h/2$. Although a more accurate prediction could be attained by using a local value of pressure in equations (\ref{segregationCoefficientSize}) and (\ref{segregationCoefficientDensity}) and solving equation (\ref{s-a-d_equation}) numerically, we show below that computations using just the mean pressure result in good agreement with the DEM simulations.

\begin{figure}[t]
\begin{center}
\includegraphics[width=0.375\columnwidth]{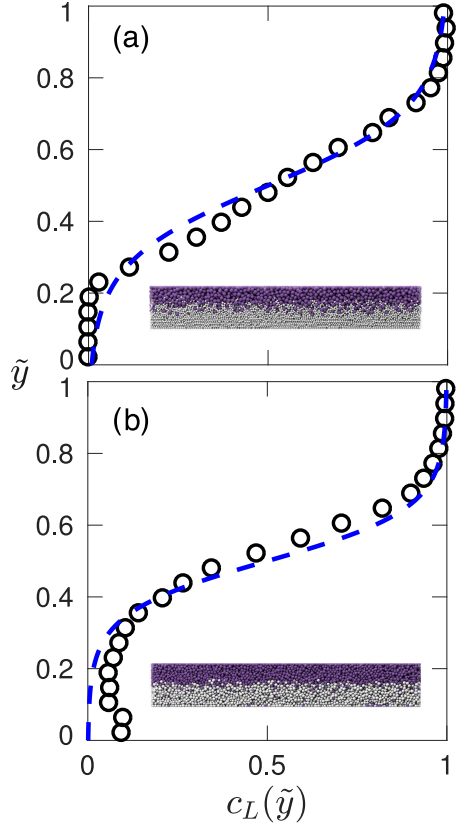}
\caption{Steady-state concentration of (a) large species, $c_{L}$, and (b) light species, $c_{L}$, vs.\ normalized height in the shear cell, $\tilde{y}=y/h$, at overburden pressure $P=\rho_B g h$ and shear rate $\dot{\gamma}=5 \, \mathrm{s^{-1}}$. Circles (o) are DEM simulation data and dashed curves (- -) are the continuum model predictions using equation (\ref{steadyStateContinuumEquation}). Insets are side view images of the DEM simulations corresponding to the concentration profiles, with dark-colored large or light particles and light-colored small or heavy particles.}\label{continuumAndDEM_singleCase}
\end{center}
\end{figure}

\begin{figure}[t]
\begin{center}
\includegraphics[width=0.375\columnwidth]{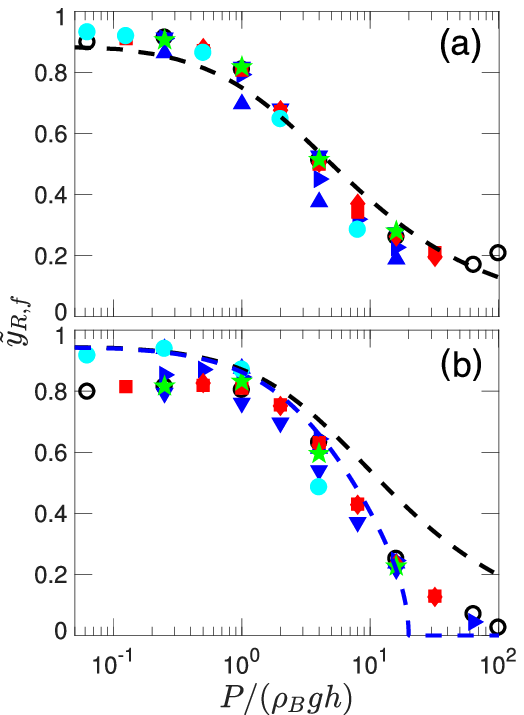}
\caption{Final rising species mean height, $\tilde{y}_{R,f}$, of (a) large species in size bidisperse flow and (b) light species in density bidisperse flow vs.\ normalized overburden pressure, $P/(\rho_B g h)$. Symbols are DEM data for the various flow conditions in Table~\ref{simulationConditions} and dashed curves (- -) are predictions of the continuum model. The blue dashed curve in (b) is calculated using a segregation cutoff at $P_{crit}=20 \rho_B g h$ and the black dashed curve is calculated without any cutoff (see text).}\label{continuumAndDEM_allCases}
\end{center}
\end{figure}

The concentration of rising species as a function of vertical location, $c_{L}(\tilde{y})$ versus\ $\tilde{y}$, for DEM results and the continuum model prediction of equation (\ref{steadyStateContinuumEquation}) are compared in Fig.~\ref{continuumAndDEM_singleCase} for a particular case at low overburden pressure of $P=\rho_B g h$. At low overburden in steady state, large or light particles in the DEM simulations are predominantly in the top half of the shear cell, with small or heavy particles concentrated in the bottom half of the shear cell. The predictions of the continuum model for size bidisperse mixtures [Fig.~\ref{continuumAndDEM_singleCase}(a)] are quantitatively similar to the DEM results other than some minor discrepancy at the lower portion of the mixing interface ($\tilde{y} \approx 0.4$) where the zones of predominantly large and small particles meet. The predictions for the density bidisperse results [Fig.~\ref{continuumAndDEM_singleCase}(b)] are also quantitatively similar except at the upper portion of the mixing interface ($\tilde{y} \approx 0.6$) and at the bottom wall ($\tilde{y} \approx 0$), where some light particles remain trapped in the DEM simulations. It is likely that the presence of a bottom wall causes light particles to be stuck in the first layer of particles in the density-bidisperse cases, since the density bidisperse particles are relatively uniform in size ($\pm 20 \%$) and thus are not easily ``lifted" off of the flat bottom wall by similarly sized particles. Nevertheless, the DEM and continuum modeling provide very similar results.

Beyond this specific case, the question is whether the simple pressure-dependent segregation rate and pressure-independent diffusion rate are generally successful at predicting steady-state segregation under various flow conditions using the continuum model. We test this by extracting a single segregation parameter from the DEM results and the analytical solution to the continuum model, denoted as $\tilde{y}_{R,f}$, that tracks the final vertical location of the center of mass of the rising species in the shear cell (e.g., large particles in size bidisperse flows or light particles in density bidisperse flows). As described previously~\cite{fry2018}, the final position of the center of mass of rising particles above the half-height of the shear cell is nondimensionalized by $h/4$ so that this metric equals 0 for perfect mixing and 1 for perfect segregation. 

We plot $\tilde{y}_{R,f}$ versus overburden pressure, $P/(\rho_B g h)$, in Fig.~\ref{continuumAndDEM_allCases} for both size and density bidisperse flows. For the size bidisperse cases in Fig.~\ref{continuumAndDEM_allCases}(a), the continuum model prediction agrees quite well with the DEM results for all of the cases tested -- the mean error between them is less than 5\% over a range of overburden pressures that corresponds to two orders of magnitude in the mean pressure in the bed. The segregation parameter, $\tilde{y}_{R,f}$, asymptotes to a value less than 1 (i.e., imperfect segregation) at low overburden, since some diffusion is always present to partially mix the particles at the interface between the two particle species. In the density bidisperse cases of Fig.~\ref{continuumAndDEM_allCases}(b), there are separate curves for two predictions based on the continuum model: one in which segregation rate is assumed to go to 0 at a critical pressure $P_{crit}=20 \rho_B g h$, above which segregation ceases altogether according to our previous study~\cite{fry2018}, and another in which a critical pressure is not included in the continuum model. Both predictions agree with the segregation data for $P<4 \rho_{B} g h$, although the continuum model tends to over-predict segregation in most cases at low overburdens by about 10 \%, since some light particles are stuck at the bottom wall of the shear cell, as noted above. At high overburdens, using the critical pressure in the continuum model predicts that the segregation ceases above the critical pressure, while the prediction without accounting for the critical pressure shows a slow decay in segregation at high overburden pressures. The DEM data displays a dependence on pressure somewhere between these two predictions. The mean error between the density bidisperse simulation results and the predictions of the continuum model that includes a critical cutoff pressure is 18\% over a range of overburden pressures  that corresponds to two orders of magnitude in the mean pressure in the bed.

\section{Enhanced mixing through pressure}\label{inverseSetup}

\begin{figure*}[ht]
\begin{center}
\includegraphics[width=1.0\textwidth]{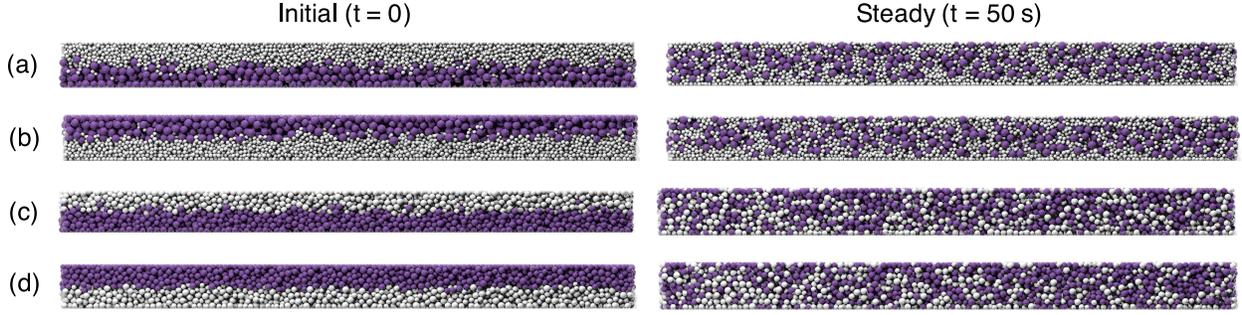}
\caption{The addition of overburden pressure suppresses segregation and leads to mixing in systems that would normally segregate. Sideview images for (a,b) size bidisperse $d_L / d_S=2$ flows with (a) smaller (light colored) particles starting above larger (dark colored) particles and (b) larger particles starting above smaller particles, and for (c,d) density bidisperse $\rho_H / \rho_L=9$ flows with (c) more dense (light-colored) particles starting above less dense (dark-colored) particles and (d) less dense particles starting above more dense particles with overburden pressure $P=20 \rho_B g h$ and a shear rate of $25 \, \mathrm{s^{-1}}$. Steady-state images are shown at $t= 50 \, \mathrm{s}$, but steady state is reached at $t \approx 10 \, \mathrm{s}$. At low pressures ($P < \rho_B g h$), these flows evolve to the condition of large (light) particles above small (heavy) particles [i.e., left images (b,d)].}\label{segregatedInitialSideShots}
\end{center}
\end{figure*}

The steady state prediction of the continuum model is insensitive to initial conditions and thus implies that shear flow under high confining pressures should mix particles that begin from an otherwise stably segregated state. To examine this hypothesis, we perform simulations of initially segregated beds at high overburden, $P=20 \rho_B g h$, above which little segregation occurs~\cite{fry2018}, corresponding to small values of $\tilde{y}_{R,f}$ in Fig.~\ref{continuumAndDEM_allCases}. Figure~\ref{segregatedInitialSideShots} shows side view images of the shear cell at the initiation of shear and after $50 \, \mathrm{s}$, at which point the system has long since reached steady state (which occurs at $t \approx 10 s$), for four cases. If low overburden pressures [i.e., $P< \rho_B g h$ in Fig.~\ref{continuumAndDEM_allCases}] were applied, all initial conditions would produce a ``stable" segregated configuration with large or light particles concentrated above small or heavy particles in the shear cell, similar to that shown for the initial conditions in Fig.~\ref{segregatedInitialSideShots}(b) for size segregation and Fig.~\ref{segregatedInitialSideShots}(d) for density segregation. However, when high confining pressure is applied, all initial conditions (left side in Fig.~\ref{segregatedInitialSideShots}) are driven to a steady state of relatively good mixing (right side in Fig.~\ref{segregatedInitialSideShots}).  What is particularly remarkable is that even in situations where the initial condition is already stably segregated [Fig.~\ref{segregatedInitialSideShots}(b) for size segregation and Fig.~\ref{segregatedInitialSideShots}(d) for density segregation], diffusion dominates because segregation is suppressed by the large overburden pressure, and both the size bidisperse and density bidisperse particles that would normally remain in their vertically graded initial positions at low overburden pressures mix due to the relatively increased impact of the diffusion. Given that confining pressure has such a demonstrably large effect on the segregation-mixing landscape, it is a useful avenue of study for applications where it is desirable to suppress segregation and enhance mixing.

\section{Discussion\label{Discussion}}
Through a series of DEM simulations of shearing granular flows, we have shown that diffusion is independent of overburden pressure in flows of equal volume mixtures of size and density bidisperse particles. The diffusion data for all simulations in which particle size, density, shear rate, and gravitational acceleration are varied collapse on a line of slope 0 when the normalized diffusion coefficient, $D_{z}/(\dot{\gamma}d^2)$, is plotted versus the normalized overburden pressure, $P/(\rho_{B} g h)$. Incorporating the pressure independence of diffusion and the pressure dependence of segregation into a continuum model for segregation previously developed for free surface flows produces results in quantitative agreement with DEM results, though further testing is needed to determine whether density segregation completely ceases at a finite critical pressure. 

Since segregation is strongly suppressed by overburden, while diffusion is nearly unaffected, the application of pressure has the potential to be used to ensure mixing in industrial applications where segregation is problematic. Experimentally testing this concept exceeds the scope of this study, but we have demonstrated computationally with a simple shear flow under confining pressure that even from initially segregated conditions, size and density bidisperse granular mixtures tend to mix in shear flows under high confining pressures. This result could yield direct improvements and new techniques in industrial areas that require precise blending. These results may also impact modeling of solid lubrication~\cite{sawyer2001,heshmat1992,wornyoh2007} and glacial till deformation~\cite{boulton1987,damsgaard2013}. At the least, the knowledge of the dependence of segregation and diffusion on overburden pressure can be incorporated into existing models of segregation in granular flows.

\acknowledgments{This research was funded by the Procter \& Gamble Company.}

\appendix*

\section{Simulation method}\label{Simulation method}
Uniform shear flows of granular materials are simulated using a soft-sphere discrete element method (DEM)~\cite{cundallStrack1979} similar to that used in our previous work~\cite{fry2018,fan2014,schlick2015,xiao2016,fan2013}. The contact forces are given by
\begin{equation}\label{normalContactEquation}
\mathbf{f_{ij}^{n}}=\big[k_{n}\mathbf{\zeta}-2\gamma_{n} m_{eff}(\mathbf{V}_{ij} \cdotp \mathbf{\hat{r}}_{ij})\big]\mathbf{\hat{r}}_{ij}
\end{equation} 
and
\begin{equation}\label{tangentialContactEquation}
\mathbf{f_{ij}^{t}}=\min{\big\{|k_{t} \beta-2\gamma_{t} m_{eff}(\mathbf{V}_{ij} \times \mathbf{\hat{r}}_{ij})|,|\mu \mathbf{F}_{ij}^{n}|\big\}}\mathrm{sgn}(\beta)\mathbf{\hat{s}}
\end{equation}
for normal and tangential contacts, respectively. In the case of static tangential contact, the tangential displacement is $\beta(t)=\int_{t_{s}}^{t} V_{ij}^{t} dt$~\cite{schaefer1996}, where $V_{ij}^{t}$ is the instantaneous tangential velocity between contacting particle surfaces, $t$ is the current time, and $t_{s}$ is the time of initial contact. For sliding tangential contact, the friction coefficient is $\mu=0.4$. The normal collision parameters are calculated as $k_{n}=\big[(\pi/t_{c})^2+\gamma_n^2\big]m_{eff}$ and $\gamma_n=-\ln{(\varepsilon)}/t_{c}$, and the tangential parameters are calculated as $k_{t}=2/7k_{n}$ and $\gamma_{t}=2/7\gamma_n$, where $\varepsilon=0.8$ is the restitution coefficient, $m_{eff}=m_{1}m_{2}/(m_{1}+m_{2})$ is the effective mass per collision, and $t_{c}=1.25 \times 10^{-4} \, \mathrm{s}$ is the binary collision time. Note that the particle stiffness $k$ changes with the inverse square of the binary collision time, $t_{c}$. When particles collide with walls, the walls are modeled as flat frictional planes and the forces are determined from the same contact equations as for particle-particle collisions.

Equation~\ref{tangentialContactEquation} contains a tangential damping term, because particles in DEM simulations under confining pressure have a tendency to oscillate about contact points in static loading, which causes nonphysical enduring kinetic energies in a particle bed that ought to be static, although the nonphysical kinetic energies are several orders of magnitude smaller than the translational kinetic energies of the particles during shear flow. The integration scheme used is the symplectic Euler algorithm, but results using the Verlet algorithm~\cite{ristow2000} display no significant differences. The symplectic Euler algorithm is also used to calculate the vertical position of the top wall, according to the balance between the wall weight and contact forces with the top layer of particles (as discussed in Section~\ref{simulationScheme}). For numerical stability, the integration timestep is $\Delta t=t_{c}/40$~\cite{schlick2015}.

\bibliography{Fry_DiffusionMixingSegregation}

\begin{thebibliography}{38}%
\makeatletter
\providecommand \@ifxundefined [1]{%
 \@ifx{#1\undefined}
}%
\providecommand \@ifnum [1]{%
 \ifnum #1\expandafter \@firstoftwo
 \else \expandafter \@secondoftwo
 \fi
}%
\providecommand \@ifx [1]{%
 \ifx #1\expandafter \@firstoftwo
 \else \expandafter \@secondoftwo
 \fi
}%
\providecommand \natexlab [1]{#1}%
\providecommand \enquote  [1]{``#1''}%
\providecommand \bibnamefont  [1]{#1}%
\providecommand \bibfnamefont [1]{#1}%
\providecommand \citenamefont [1]{#1}%
\providecommand \href@noop [0]{\@secondoftwo}%
\providecommand \href [0]{\begingroup \@sanitize@url \@href}%
\providecommand \@href[1]{\@@startlink{#1}\@@href}%
\providecommand \@@href[1]{\endgroup#1\@@endlink}%
\providecommand \@sanitize@url [0]{\catcode `\\12\catcode `\$12\catcode
  `\&12\catcode `\#12\catcode `\^12\catcode `\_12\catcode `\%12\relax}%
\providecommand \@@startlink[1]{}%
\providecommand \@@endlink[0]{}%
\providecommand \url  [0]{\begingroup\@sanitize@url \@url }%
\providecommand \@url [1]{\endgroup\@href {#1}{\urlprefix }}%
\providecommand \urlprefix  [0]{URL }%
\providecommand \Eprint [0]{\href }%
\providecommand \doibase [0]{http://dx.doi.org/}%
\providecommand \selectlanguage [0]{\@gobble}%
\providecommand \bibinfo  [0]{\@secondoftwo}%
\providecommand \bibfield  [0]{\@secondoftwo}%
\providecommand \translation [1]{[#1]}%
\providecommand \BibitemOpen [0]{}%
\providecommand \bibitemStop [0]{}%
\providecommand \bibitemNoStop [0]{.\EOS\space}%
\providecommand \EOS [0]{\spacefactor3000\relax}%
\providecommand \BibitemShut  [1]{\csname bibitem#1\endcsname}%
\let\auto@bib@innerbib\@empty
\bibitem [{\citenamefont {Savage}\ and\ \citenamefont
  {Lun}(1988)}]{savageLun1988}%
  \BibitemOpen
  \bibfield  {author} {\bibinfo {author} {\bibfnamefont {S.~B.}\ \bibnamefont
  {Savage}}\ and\ \bibinfo {author} {\bibfnamefont {C.~K.~K.}\ \bibnamefont
  {Lun}},\ }\href@noop {} {\bibfield  {journal} {\bibinfo  {journal} {J. Fluid
  Mech.}\ }\textbf {\bibinfo {volume} {189}},\ \bibinfo {pages} {311} (\bibinfo
  {year} {1988})}\BibitemShut {NoStop}%
\bibitem [{\citenamefont {Ottino}\ and\ \citenamefont
  {Khakhar}(2000)}]{ottinoKhakhar2000}%
  \BibitemOpen
  \bibfield  {author} {\bibinfo {author} {\bibfnamefont {J.~M.}\ \bibnamefont
  {Ottino}}\ and\ \bibinfo {author} {\bibfnamefont {D.~V.}\ \bibnamefont
  {Khakhar}},\ }\href@noop {} {\bibfield  {journal} {\bibinfo  {journal} {Annu.
  Rev. Fluid Mech.}\ }\textbf {\bibinfo {volume} {32}},\ \bibinfo {pages} {55}
  (\bibinfo {year} {2000})}\BibitemShut {NoStop}%
\bibitem [{\citenamefont {Dolgunin}\ and\ \citenamefont
  {Ukolov}(1995)}]{dolgunin1995}%
  \BibitemOpen
  \bibfield  {author} {\bibinfo {author} {\bibfnamefont {V.~N.}\ \bibnamefont
  {Dolgunin}}\ and\ \bibinfo {author} {\bibfnamefont {A.~A.}\ \bibnamefont
  {Ukolov}},\ }\href@noop {} {\bibfield  {journal} {\bibinfo  {journal} {Powder
  Tech.}\ }\textbf {\bibinfo {volume} {83}},\ \bibinfo {pages} {95} (\bibinfo
  {year} {1995})}\BibitemShut {NoStop}%
\bibitem [{\citenamefont {Gray}\ and\ \citenamefont
  {Thornton}(2005)}]{grayThornton2005}%
  \BibitemOpen
  \bibfield  {author} {\bibinfo {author} {\bibfnamefont {J.~M. N.~T.}\
  \bibnamefont {Gray}}\ and\ \bibinfo {author} {\bibfnamefont {A.~R.}\
  \bibnamefont {Thornton}},\ }\href@noop {} {\bibfield  {journal} {\bibinfo
  {journal} {Proc. R. Soc. A}\ }\textbf {\bibinfo {volume} {461}},\ \bibinfo
  {pages} {1447} (\bibinfo {year} {2005})}\BibitemShut {NoStop}%
\bibitem [{\citenamefont {Fan}\ \emph {et~al.}(2014)\citenamefont {Fan},
  \citenamefont {Schlick}, \citenamefont {Umbanhowar}, \citenamefont {Ottino},\
  and\ \citenamefont {Lueptow}}]{fan2014}%
  \BibitemOpen
  \bibfield  {author} {\bibinfo {author} {\bibfnamefont {Y.}~\bibnamefont
  {Fan}}, \bibinfo {author} {\bibfnamefont {C.~P.}\ \bibnamefont {Schlick}},
  \bibinfo {author} {\bibfnamefont {P.~B.}\ \bibnamefont {Umbanhowar}},
  \bibinfo {author} {\bibfnamefont {J.~M.}\ \bibnamefont {Ottino}}, \ and\
  \bibinfo {author} {\bibfnamefont {R.~M.}\ \bibnamefont {Lueptow}},\
  }\href@noop {} {\bibfield  {journal} {\bibinfo  {journal} {J. Fluid Mech.}\
  }\textbf {\bibinfo {volume} {741}},\ \bibinfo {pages} {252} (\bibinfo {year}
  {2014})}\BibitemShut {NoStop}%
\bibitem [{\citenamefont {Bridgwater}(1980)}]{bridgwater1980}%
  \BibitemOpen
  \bibfield  {author} {\bibinfo {author} {\bibfnamefont {J.}~\bibnamefont
  {Bridgwater}},\ }\href@noop {} {\bibfield  {journal} {\bibinfo  {journal}
  {Powder Tech.}\ }\textbf {\bibinfo {volume} {25}},\ \bibinfo {pages} {129}
  (\bibinfo {year} {1980})}\BibitemShut {NoStop}%
\bibitem [{\citenamefont {Utter}\ and\ \citenamefont
  {Behringer}(2004)}]{utter2004}%
  \BibitemOpen
  \bibfield  {author} {\bibinfo {author} {\bibfnamefont {B.}~\bibnamefont
  {Utter}}\ and\ \bibinfo {author} {\bibfnamefont {R.~P.}\ \bibnamefont
  {Behringer}},\ }\href@noop {} {\bibfield  {journal} {\bibinfo  {journal}
  {Phys. Rev. E}\ }\textbf {\bibinfo {volume} {69}},\ \bibinfo {pages} {031308}
  (\bibinfo {year} {2004})}\BibitemShut {NoStop}%
\bibitem [{\citenamefont {Drahun}\ and\ \citenamefont
  {Bridgwater}(1983)}]{drahunBridgwater1983}%
  \BibitemOpen
  \bibfield  {author} {\bibinfo {author} {\bibfnamefont {J.~A.}\ \bibnamefont
  {Drahun}}\ and\ \bibinfo {author} {\bibfnamefont {J.}~\bibnamefont
  {Bridgwater}},\ }\href@noop {} {\bibfield  {journal} {\bibinfo  {journal}
  {Powder Tech.}\ }\textbf {\bibinfo {volume} {36}},\ \bibinfo {pages} {39}
  (\bibinfo {year} {1983})}\BibitemShut {NoStop}%
\bibitem [{\citenamefont {Schlick}\ \emph
  {et~al.}(2015{\natexlab{a}})\citenamefont {Schlick}, \citenamefont {Fan},
  \citenamefont {Isner}, \citenamefont {Umbanhowar}, \citenamefont {Ottino},\
  and\ \citenamefont {Lueptow}}]{schlick2015}%
  \BibitemOpen
  \bibfield  {author} {\bibinfo {author} {\bibfnamefont {C.~P.}\ \bibnamefont
  {Schlick}}, \bibinfo {author} {\bibfnamefont {Y.}~\bibnamefont {Fan}},
  \bibinfo {author} {\bibfnamefont {A.~B.}\ \bibnamefont {Isner}}, \bibinfo
  {author} {\bibfnamefont {P.~B.}\ \bibnamefont {Umbanhowar}}, \bibinfo
  {author} {\bibfnamefont {J.~M.}\ \bibnamefont {Ottino}}, \ and\ \bibinfo
  {author} {\bibfnamefont {R.~M.}\ \bibnamefont {Lueptow}},\ }\href@noop {}
  {\bibfield  {journal} {\bibinfo  {journal} {AIChE J.}\ }\textbf {\bibinfo
  {volume} {61}},\ \bibinfo {pages} {1524} (\bibinfo {year}
  {2015}{\natexlab{a}})}\BibitemShut {NoStop}%
\bibitem [{\citenamefont {Tripathi}\ and\ \citenamefont
  {Khakhar}(2013)}]{tripathi2013}%
  \BibitemOpen
  \bibfield  {author} {\bibinfo {author} {\bibfnamefont {A.}~\bibnamefont
  {Tripathi}}\ and\ \bibinfo {author} {\bibfnamefont {D.~V.}\ \bibnamefont
  {Khakhar}},\ }\href@noop {} {\bibfield  {journal} {\bibinfo  {journal} {J.
  Fluid Mech.}\ }\textbf {\bibinfo {volume} {717}},\ \bibinfo {pages} {643}
  (\bibinfo {year} {2013})}\BibitemShut {NoStop}%
\bibitem [{\citenamefont {Xiao}\ \emph {et~al.}(2016)\citenamefont {Xiao},
  \citenamefont {Umbanhowar}, \citenamefont {Ottino},\ and\ \citenamefont
  {Lueptow}}]{xiao2016}%
  \BibitemOpen
  \bibfield  {author} {\bibinfo {author} {\bibfnamefont {H.}~\bibnamefont
  {Xiao}}, \bibinfo {author} {\bibfnamefont {P.~B.}\ \bibnamefont
  {Umbanhowar}}, \bibinfo {author} {\bibfnamefont {J.~M.}\ \bibnamefont
  {Ottino}}, \ and\ \bibinfo {author} {\bibfnamefont {R.~M.}\ \bibnamefont
  {Lueptow}},\ }\href@noop {} {\bibfield  {journal} {\bibinfo  {journal} {Proc.
  R. Soc. A}\ }\textbf {\bibinfo {volume} {472}},\ \bibinfo {pages} {20150856}
  (\bibinfo {year} {2016})}\BibitemShut {NoStop}%
\bibitem [{\citenamefont {Liu}\ and\ \citenamefont {McCarthy}(2017)}]{liu2017}%
  \BibitemOpen
  \bibfield  {author} {\bibinfo {author} {\bibfnamefont {S.}~\bibnamefont
  {Liu}}\ and\ \bibinfo {author} {\bibfnamefont {J.~J.}\ \bibnamefont
  {McCarthy}},\ }\href@noop {} {\bibfield  {journal} {\bibinfo  {journal}
  {Phys. Rev. E}\ }\textbf {\bibinfo {volume} {96}},\ \bibinfo {pages} {020901}
  (\bibinfo {year} {2017})}\BibitemShut {NoStop}%
\bibitem [{\citenamefont {Gray}\ and\ \citenamefont
  {Chugunov}(2006)}]{gray2006}%
  \BibitemOpen
  \bibfield  {author} {\bibinfo {author} {\bibfnamefont {J.~M. N.~T.}\
  \bibnamefont {Gray}}\ and\ \bibinfo {author} {\bibfnamefont {V.~A.}\
  \bibnamefont {Chugunov}},\ }\href@noop {} {\bibfield  {journal} {\bibinfo
  {journal} {J. Fluid Mech.}\ }\textbf {\bibinfo {volume} {569}},\ \bibinfo
  {pages} {365} (\bibinfo {year} {2006})}\BibitemShut {NoStop}%
\bibitem [{\citenamefont {Hill}\ and\ \citenamefont {Fan}(2008)}]{hillFan2008}%
  \BibitemOpen
  \bibfield  {author} {\bibinfo {author} {\bibfnamefont {K.~M.}\ \bibnamefont
  {Hill}}\ and\ \bibinfo {author} {\bibfnamefont {Y.}~\bibnamefont {Fan}},\
  }\href@noop {} {\bibfield  {journal} {\bibinfo  {journal} {Phys. Rev. Lett.}\
  }\textbf {\bibinfo {volume} {101}},\ \bibinfo {pages} {088001} (\bibinfo
  {year} {2008})}\BibitemShut {NoStop}%
\bibitem [{\citenamefont {Golick}\ and\ \citenamefont
  {Daniels}(2009)}]{golickDaniels2009}%
  \BibitemOpen
  \bibfield  {author} {\bibinfo {author} {\bibfnamefont {L.~A.}\ \bibnamefont
  {Golick}}\ and\ \bibinfo {author} {\bibfnamefont {K.~E.}\ \bibnamefont
  {Daniels}},\ }\href@noop {} {\bibfield  {journal} {\bibinfo  {journal} {Phys.
  Rev. E}\ }\textbf {\bibinfo {volume} {80}},\ \bibinfo {pages} {042301}
  (\bibinfo {year} {2009})}\BibitemShut {NoStop}%
\bibitem [{\citenamefont {Bridgwater}(1994)}]{bridgwater1994}%
  \BibitemOpen
  \bibfield  {author} {\bibinfo {author} {\bibfnamefont {J.}~\bibnamefont
  {Bridgwater}},\ }in\ \href@noop {} {\emph {\bibinfo {booktitle} {Granular
  Matter}}}\ (\bibinfo  {publisher} {Springer},\ \bibinfo {year} {1994})\ pp.\
  \bibinfo {pages} {161--193}\BibitemShut {NoStop}%
\bibitem [{\citenamefont {Khola}\ and\ \citenamefont
  {Wassgren}(2016)}]{khola2016}%
  \BibitemOpen
  \bibfield  {author} {\bibinfo {author} {\bibfnamefont {N.}~\bibnamefont
  {Khola}}\ and\ \bibinfo {author} {\bibfnamefont {C.}~\bibnamefont
  {Wassgren}},\ }\href@noop {} {\bibfield  {journal} {\bibinfo  {journal}
  {Powder Tech.}\ }\textbf {\bibinfo {volume} {288}},\ \bibinfo {pages} {441}
  (\bibinfo {year} {2016})}\BibitemShut {NoStop}%
\bibitem [{\citenamefont {Fry}\ \emph {et~al.}(2018)\citenamefont {Fry},
  \citenamefont {Umbanhowar}, \citenamefont {Ottino},\ and\ \citenamefont
  {Lueptow}}]{fry2018}%
  \BibitemOpen
  \bibfield  {author} {\bibinfo {author} {\bibfnamefont {A.~M.}\ \bibnamefont
  {Fry}}, \bibinfo {author} {\bibfnamefont {P.~B.}\ \bibnamefont {Umbanhowar}},
  \bibinfo {author} {\bibfnamefont {J.~M.}\ \bibnamefont {Ottino}}, \ and\
  \bibinfo {author} {\bibfnamefont {R.~M.}\ \bibnamefont {Lueptow}},\ }\href
  {\doibase 10.1103/PhysRevE.97.062906} {\bibfield  {journal} {\bibinfo
  {journal} {Phys. Rev. E}\ }\textbf {\bibinfo {volume} {97}},\ \bibinfo
  {pages} {062906} (\bibinfo {year} {2018})}\BibitemShut {NoStop}%
\bibitem [{\citenamefont {Gray}\ and\ \citenamefont
  {Ancey}(2015)}]{grayAncey2015}%
  \BibitemOpen
  \bibfield  {author} {\bibinfo {author} {\bibfnamefont {J.~M. N.~T.}\
  \bibnamefont {Gray}}\ and\ \bibinfo {author} {\bibfnamefont {C.}~\bibnamefont
  {Ancey}},\ }\href@noop {} {\bibfield  {journal} {\bibinfo  {journal} {J.
  Fluid Mech.}\ }\textbf {\bibinfo {volume} {779}},\ \bibinfo {pages} {622}
  (\bibinfo {year} {2015})}\BibitemShut {NoStop}%
\bibitem [{\citenamefont {May}\ \emph {et~al.}(2010)\citenamefont {May},
  \citenamefont {Golick}, \citenamefont {Phillips}, \citenamefont {Shearer},\
  and\ \citenamefont {Daniels}}]{mayDaniels2010}%
  \BibitemOpen
  \bibfield  {author} {\bibinfo {author} {\bibfnamefont {L.~B.~H.}\
  \bibnamefont {May}}, \bibinfo {author} {\bibfnamefont {L.~A.}\ \bibnamefont
  {Golick}}, \bibinfo {author} {\bibfnamefont {K.~C.}\ \bibnamefont
  {Phillips}}, \bibinfo {author} {\bibfnamefont {M.}~\bibnamefont {Shearer}}, \
  and\ \bibinfo {author} {\bibfnamefont {K.~E.}\ \bibnamefont {Daniels}},\
  }\href@noop {} {\bibfield  {journal} {\bibinfo  {journal} {Phys. Rev. E}\
  }\textbf {\bibinfo {volume} {81}},\ \bibinfo {pages} {051301} (\bibinfo
  {year} {2010})}\BibitemShut {NoStop}%
\bibitem [{\citenamefont {Lerner}\ \emph {et~al.}(2012)\citenamefont {Lerner},
  \citenamefont {D{\"u}ring},\ and\ \citenamefont {Wyart}}]{lerner2012}%
  \BibitemOpen
  \bibfield  {author} {\bibinfo {author} {\bibfnamefont {E.}~\bibnamefont
  {Lerner}}, \bibinfo {author} {\bibfnamefont {G.}~\bibnamefont {D{\"u}ring}},
  \ and\ \bibinfo {author} {\bibfnamefont {M.}~\bibnamefont {Wyart}},\
  }\href@noop {} {\bibfield  {journal} {\bibinfo  {journal} {Proc. Natl. Acad.
  Sci. U.S.A.}\ }\textbf {\bibinfo {volume} {109}},\ \bibinfo {pages} {4798}
  (\bibinfo {year} {2012})}\BibitemShut {NoStop}%
\bibitem [{\citenamefont {Clark}\ \emph {et~al.}(2018)\citenamefont {Clark},
  \citenamefont {Thompson}, \citenamefont {Shattuck}, \citenamefont
  {Ouellette},\ and\ \citenamefont {O'Hern}}]{clark2018}%
  \BibitemOpen
  \bibfield  {author} {\bibinfo {author} {\bibfnamefont {A.~H.}\ \bibnamefont
  {Clark}}, \bibinfo {author} {\bibfnamefont {J.~D.}\ \bibnamefont {Thompson}},
  \bibinfo {author} {\bibfnamefont {M.~D.}\ \bibnamefont {Shattuck}}, \bibinfo
  {author} {\bibfnamefont {N.~T.}\ \bibnamefont {Ouellette}}, \ and\ \bibinfo
  {author} {\bibfnamefont {C.~S.}\ \bibnamefont {O'Hern}},\ }\href@noop {}
  {\bibfield  {journal} {\bibinfo  {journal} {Phys. Rev. E}\ }\textbf {\bibinfo
  {volume} {97}},\ \bibinfo {pages} {062901} (\bibinfo {year}
  {2018})}\BibitemShut {NoStop}%
\bibitem [{\citenamefont {Savage}\ and\ \citenamefont
  {Dai}(1993)}]{savage1993}%
  \BibitemOpen
  \bibfield  {author} {\bibinfo {author} {\bibfnamefont {S.~B.}\ \bibnamefont
  {Savage}}\ and\ \bibinfo {author} {\bibfnamefont {R.}~\bibnamefont {Dai}},\
  }\href@noop {} {\bibfield  {journal} {\bibinfo  {journal} {Mech. Materials}\
  }\textbf {\bibinfo {volume} {16}},\ \bibinfo {pages} {225} (\bibinfo {year}
  {1993})}\BibitemShut {NoStop}%
\bibitem [{\citenamefont {Campbell}(1997)}]{campbell1997}%
  \BibitemOpen
  \bibfield  {author} {\bibinfo {author} {\bibfnamefont {C.~S.}\ \bibnamefont
  {Campbell}},\ }\href@noop {} {\bibfield  {journal} {\bibinfo  {journal} {J.
  Fluid Mech.}\ }\textbf {\bibinfo {volume} {348}},\ \bibinfo {pages} {85}
  (\bibinfo {year} {1997})}\BibitemShut {NoStop}%
\bibitem [{\citenamefont {Hsiau}\ and\ \citenamefont
  {Shieh}(1999)}]{hsiau1999}%
  \BibitemOpen
  \bibfield  {author} {\bibinfo {author} {\bibfnamefont {S.~S.}\ \bibnamefont
  {Hsiau}}\ and\ \bibinfo {author} {\bibfnamefont {Y.~M.}\ \bibnamefont
  {Shieh}},\ }\href@noop {} {\bibfield  {journal} {\bibinfo  {journal} {J.
  Rheology}\ }\textbf {\bibinfo {volume} {43}},\ \bibinfo {pages} {1049}
  (\bibinfo {year} {1999})}\BibitemShut {NoStop}%
\bibitem [{\citenamefont {Fan}\ \emph {et~al.}(2013)\citenamefont {Fan},
  \citenamefont {Umbanhowar}, \citenamefont {Ottino},\ and\ \citenamefont
  {Lueptow}}]{fan2013}%
  \BibitemOpen
  \bibfield  {author} {\bibinfo {author} {\bibfnamefont {Y.}~\bibnamefont
  {Fan}}, \bibinfo {author} {\bibfnamefont {P.~B.}\ \bibnamefont {Umbanhowar}},
  \bibinfo {author} {\bibfnamefont {J.~M.}\ \bibnamefont {Ottino}}, \ and\
  \bibinfo {author} {\bibfnamefont {R.~M.}\ \bibnamefont {Lueptow}},\
  }\href@noop {} {\bibfield  {journal} {\bibinfo  {journal} {Proc. R. Soc. A}\
  }\textbf {\bibinfo {volume} {469}},\ \bibinfo {pages} {20130235} (\bibinfo
  {year} {2013})}\BibitemShut {NoStop}%
\bibitem [{\citenamefont {Fan}\ \emph {et~al.}(2015)\citenamefont {Fan},
  \citenamefont {Umbanhowar}, \citenamefont {Ottino},\ and\ \citenamefont
  {Lueptow}}]{fan2015}%
  \BibitemOpen
  \bibfield  {author} {\bibinfo {author} {\bibfnamefont {Y.}~\bibnamefont
  {Fan}}, \bibinfo {author} {\bibfnamefont {P.~B.}\ \bibnamefont {Umbanhowar}},
  \bibinfo {author} {\bibfnamefont {J.~M.}\ \bibnamefont {Ottino}}, \ and\
  \bibinfo {author} {\bibfnamefont {R.~M.}\ \bibnamefont {Lueptow}},\
  }\href@noop {} {\bibfield  {journal} {\bibinfo  {journal} {Phys. Rev. Lett.}\
  }\textbf {\bibinfo {volume} {115}},\ \bibinfo {pages} {088001} (\bibinfo
  {year} {2015})}\BibitemShut {NoStop}%
\bibitem [{\citenamefont {van~der Vaart}\ \emph {et~al.}(2015)\citenamefont
  {van~der Vaart}, \citenamefont {Gajjar}, \citenamefont {Epely-Chauvin},
  \citenamefont {Andreini}, \citenamefont {Gray},\ and\ \citenamefont
  {Ancey}}]{vanderVaart2015}%
  \BibitemOpen
  \bibfield  {author} {\bibinfo {author} {\bibfnamefont {K.}~\bibnamefont
  {van~der Vaart}}, \bibinfo {author} {\bibfnamefont {P.}~\bibnamefont
  {Gajjar}}, \bibinfo {author} {\bibfnamefont {G.}~\bibnamefont
  {Epely-Chauvin}}, \bibinfo {author} {\bibfnamefont {N.}~\bibnamefont
  {Andreini}}, \bibinfo {author} {\bibfnamefont {J.~M. N.~T.}\ \bibnamefont
  {Gray}}, \ and\ \bibinfo {author} {\bibfnamefont {C.}~\bibnamefont {Ancey}},\
  }\href@noop {} {\bibfield  {journal} {\bibinfo  {journal} {Phys. Rev. Lett.}\
  }\textbf {\bibinfo {volume} {114}},\ \bibinfo {pages} {238001} (\bibinfo
  {year} {2015})}\BibitemShut {NoStop}%
\bibitem [{\citenamefont {Jones}\ \emph {et~al.}(2018)\citenamefont {Jones},
  \citenamefont {Isner}, \citenamefont {Xiao}, \citenamefont {Ottino},
  \citenamefont {Umbanhowar},\ and\ \citenamefont {Lueptow}}]{jones2018}%
  \BibitemOpen
  \bibfield  {author} {\bibinfo {author} {\bibfnamefont {R.~P.}\ \bibnamefont
  {Jones}}, \bibinfo {author} {\bibfnamefont {A.~B.}\ \bibnamefont {Isner}},
  \bibinfo {author} {\bibfnamefont {H.}~\bibnamefont {Xiao}}, \bibinfo {author}
  {\bibfnamefont {J.~M.}\ \bibnamefont {Ottino}}, \bibinfo {author}
  {\bibfnamefont {P.~B.}\ \bibnamefont {Umbanhowar}}, \ and\ \bibinfo {author}
  {\bibfnamefont {R.~M.}\ \bibnamefont {Lueptow}},\ }\href@noop {} {\bibfield
  {journal} {\bibinfo  {journal} {Phys. Rev. Fluids, to appear}\ } (\bibinfo
  {year} {2018})}\BibitemShut {NoStop}%
\bibitem [{\citenamefont {Schlick}\ \emph
  {et~al.}(2015{\natexlab{b}})\citenamefont {Schlick}, \citenamefont {Isner},
  \citenamefont {Umbanhowar}, \citenamefont {Lueptow},\ and\ \citenamefont
  {Ottino}}]{schlick2015b}%
  \BibitemOpen
  \bibfield  {author} {\bibinfo {author} {\bibfnamefont {C.~P.}\ \bibnamefont
  {Schlick}}, \bibinfo {author} {\bibfnamefont {A.~B.}\ \bibnamefont {Isner}},
  \bibinfo {author} {\bibfnamefont {P.~B.}\ \bibnamefont {Umbanhowar}},
  \bibinfo {author} {\bibfnamefont {R.~M.}\ \bibnamefont {Lueptow}}, \ and\
  \bibinfo {author} {\bibfnamefont {J.~M.}\ \bibnamefont {Ottino}},\
  }\href@noop {} {\bibfield  {journal} {\bibinfo  {journal} {Industrial \&
  Engineering Chemistry Research}\ }\textbf {\bibinfo {volume} {54}},\ \bibinfo
  {pages} {10465} (\bibinfo {year} {2015}{\natexlab{b}})}\BibitemShut {NoStop}%
\bibitem [{\citenamefont {Sawyer}\ and\ \citenamefont
  {Tichy}(2001)}]{sawyer2001}%
  \BibitemOpen
  \bibfield  {author} {\bibinfo {author} {\bibfnamefont {W.~G.}\ \bibnamefont
  {Sawyer}}\ and\ \bibinfo {author} {\bibfnamefont {J.~A.}\ \bibnamefont
  {Tichy}},\ }\href@noop {} {\bibfield  {journal} {\bibinfo  {journal} {J.
  Tribology}\ }\textbf {\bibinfo {volume} {123}},\ \bibinfo {pages} {777}
  (\bibinfo {year} {2001})}\BibitemShut {NoStop}%
\bibitem [{\citenamefont {Heshmat}(1992)}]{heshmat1992}%
  \BibitemOpen
  \bibfield  {author} {\bibinfo {author} {\bibfnamefont {H.}~\bibnamefont
  {Heshmat}},\ }\href@noop {} {\bibfield  {journal} {\bibinfo  {journal}
  {Lubrication Eng.}\ }\textbf {\bibinfo {volume} {48}},\ \bibinfo {pages}
  {373} (\bibinfo {year} {1992})}\BibitemShut {NoStop}%
\bibitem [{\citenamefont {Wornyoh}\ \emph {et~al.}(2007)\citenamefont
  {Wornyoh}, \citenamefont {Jasti},\ and\ \citenamefont {Higgs}}]{wornyoh2007}%
  \BibitemOpen
  \bibfield  {author} {\bibinfo {author} {\bibfnamefont {E.~Y.~A.}\
  \bibnamefont {Wornyoh}}, \bibinfo {author} {\bibfnamefont {V.~K.}\
  \bibnamefont {Jasti}}, \ and\ \bibinfo {author} {\bibfnamefont {C.~F.}\
  \bibnamefont {Higgs}},\ }\href@noop {} {\bibfield  {journal} {\bibinfo
  {journal} {J. Tribology}\ }\textbf {\bibinfo {volume} {129}},\ \bibinfo
  {pages} {438} (\bibinfo {year} {2007})}\BibitemShut {NoStop}%
\bibitem [{\citenamefont {Boulton}\ and\ \citenamefont
  {Hindmarsh}(1987)}]{boulton1987}%
  \BibitemOpen
  \bibfield  {author} {\bibinfo {author} {\bibfnamefont {G.}~\bibnamefont
  {Boulton}}\ and\ \bibinfo {author} {\bibfnamefont {R.}~\bibnamefont
  {Hindmarsh}},\ }\href@noop {} {\bibfield  {journal} {\bibinfo  {journal} {J.
  Geophys. Res.}\ }\textbf {\bibinfo {volume} {92}},\ \bibinfo {pages} {9059}
  (\bibinfo {year} {1987})}\BibitemShut {NoStop}%
\bibitem [{\citenamefont {Damsgaard}\ \emph {et~al.}(2013)\citenamefont
  {Damsgaard}, \citenamefont {Egholm}, \citenamefont {Piotrowski},
  \citenamefont {Tulaczyk}, \citenamefont {Larsen},\ and\ \citenamefont
  {Tylmann}}]{damsgaard2013}%
  \BibitemOpen
  \bibfield  {author} {\bibinfo {author} {\bibfnamefont {A.}~\bibnamefont
  {Damsgaard}}, \bibinfo {author} {\bibfnamefont {D.~L.}\ \bibnamefont
  {Egholm}}, \bibinfo {author} {\bibfnamefont {J.~A.}\ \bibnamefont
  {Piotrowski}}, \bibinfo {author} {\bibfnamefont {S.}~\bibnamefont
  {Tulaczyk}}, \bibinfo {author} {\bibfnamefont {N.~K.}\ \bibnamefont
  {Larsen}}, \ and\ \bibinfo {author} {\bibfnamefont {K.}~\bibnamefont
  {Tylmann}},\ }\href@noop {} {\bibfield  {journal} {\bibinfo  {journal} {J.
  Geophys. Res.: Earth Surface}\ }\textbf {\bibinfo {volume} {118}},\ \bibinfo
  {pages} {2230} (\bibinfo {year} {2013})}\BibitemShut {NoStop}%
\bibitem [{\citenamefont {Cundall}\ and\ \citenamefont
  {Strack}(1979)}]{cundallStrack1979}%
  \BibitemOpen
  \bibfield  {author} {\bibinfo {author} {\bibfnamefont {P.~A.}\ \bibnamefont
  {Cundall}}\ and\ \bibinfo {author} {\bibfnamefont {O.~D.~L.}\ \bibnamefont
  {Strack}},\ }\href@noop {} {\bibfield  {journal} {\bibinfo  {journal}
  {G{\'e}otechnique}\ }\textbf {\bibinfo {volume} {29}},\ \bibinfo {pages} {47}
  (\bibinfo {year} {1979})}\BibitemShut {NoStop}%
\bibitem [{\citenamefont {Sch{\"a}fer}\ \emph {et~al.}(1996)\citenamefont
  {Sch{\"a}fer}, \citenamefont {Dippel},\ and\ \citenamefont
  {Wolf}}]{schaefer1996}%
  \BibitemOpen
  \bibfield  {author} {\bibinfo {author} {\bibfnamefont {J.}~\bibnamefont
  {Sch{\"a}fer}}, \bibinfo {author} {\bibfnamefont {S.}~\bibnamefont {Dippel}},
  \ and\ \bibinfo {author} {\bibfnamefont {D.~E.}\ \bibnamefont {Wolf}},\
  }\href@noop {} {\bibfield  {journal} {\bibinfo  {journal} {J. Phys. I}\
  }\textbf {\bibinfo {volume} {6}},\ \bibinfo {pages} {5} (\bibinfo {year}
  {1996})}\BibitemShut {NoStop}%
\bibitem [{\citenamefont {Ristow}(2000)}]{ristow2000}%
  \BibitemOpen
  \bibfield  {author} {\bibinfo {author} {\bibfnamefont {G.~H.}\ \bibnamefont
  {Ristow}},\ }\href@noop {} {\emph {\bibinfo {title} {Pattern Formation in
  Granular Materials}}},\ \bibinfo {series} {Springer Tracts in Modern
  Physics}, Vol.\ \bibinfo {volume} {164}\ (\bibinfo  {publisher} {Springer,
  Berlin},\ \bibinfo {year} {2000})\BibitemShut {NoStop}%
\end{thebibliography}%
\bibliographystyle{apsrev4-1}

\end{document}